\documentstyle[11pt]{article}
\begin{document}
\begin{center}
\large{
A Symmetry-induced Model of Elliptical Galaxy Patterns
}
\normalsize\\
Jin He\\
Department of Physics and Astronomy, the University of Alabama, Tuscaloosa, AL35487\\
E-mail:jin002@bama.ua.edu\\
\mbox{   }
\end{center}
\large{{\bf Abstract} } \normalsize 
S\'ersic (1968) generalized the de Vaucouleurs law which follows the projected (observed) one dimensional radial profile of elliptical galaxies  closely and 
Dehnen (1993) proposed an analytical formula of the 3-dimensional light distributions whose projected line profile resembles the de Vaucouleurs law.
This paper is involved to recover the Dehnen model and generalize the model to account for galaxy elliptical shapes by means of curvilinear coordinate systems and employing a symmetry principle. The symmetry principle maps an orthogonal coordinate system to a light distribution pattern.
The coordinate system for elliptical galaxy patterns turns out to be the one which is formed by the complex-plane reciprocal transformation $Z=1/W$. 
The resulting spatial (3-dimensional) light distribution is spherically symmetric and has infinite gradient at its centre, which is called spherical-nucleus solution and is used to model galaxy central area. We can make changes of the coordinate system by cutting out some column areas of its definition domain, the areas containing the galaxy centre.   The resulting spatial (3-dimensional) light distributions are axisymmetric or triaxial and have zero gradient at the centre, which are called elliptical-shape solutions and are used to model global elliptical patterns. 
The two types of logarithmic light distributions are added together to model full elliptical galaxy patterns. The model is a generalization of the Dehnen model.     
One of the elliptical-shape solutions permits realistic numerical calculation and 
is fitted to all R-band elliptical images from the Frei {\it et al. }(1996)'s galaxy sample. The fitting is satisfactory.
This suggests that elliptical galaxy patterns can be represented in terms of a few basic parameters. \\
keywords \,\, Methods : Analytical -- Galaxies : Structure\\
\\
\section{Introduction}

Compared to disc galaxies, elliptical galaxies are more ``clean'' objects for people to test models on the galactic-scale gravitational systems because they have smoother texture of the photographic images. Their light distributions can be approximately the linear representation of their mass distributions.
Real elliptical galaxies are 3-dimensional distributions of light.
The observed 2-dimensional elliptical-shaped light distributions (i.e., surface brightness) result from the integration of the light along the line-of-sight, i.\,e., the projection on to the 2-dimensional sky plane.
Therefore, a complete understanding of elliptical galaxy patterns is the analytical formulation of their 3-dimensional light distributions, i.\,e., the analytical deprojection of the observed 2-dimensional profiles. 
The 3-dimensional light distributions have two aspects: one is the law of their one dimensional radial density profile and the other is their 3-dimensional shapes. 

Fortunately, the projected one dimensional radial profile is studied very clear.
It decreases smoothly from the nucleus, closely following the S\'ersic law  (1968)
\begin{equation}
\Sigma _s(R)=\sigma _s\exp(\sigma _0 R^{1/n})
\end{equation}
where $\sigma _s (>0)$, $\sigma _0 (<0)$ are constants and  $R$ is the projected radial distance from the galaxy centre along the semi-major axis (however, the spatial (3-dimensional) radial distance from the galaxy centre is denoted, from now on, by $r$).
The S\'ersic law, which fits galaxy inner parts very well, indicates that galaxy light gradients at their centres are $ - \infty$.
The Hubble (1930) law suggests an unrealistic zero-gradient at the galaxy centres (flattened light distribution). 
Much attention has been given to the deprojection of S\'ersic law, that is, the proposition of the spatial (3-dimensional) light distributions (or equivalently, in the case of spherically symmetric light distributions, the proposition of the spatial one dimensional radial  profile) whose projected line profiles follow the S\'ersic law.  
However, the deprojection of the S\'ersic law is not tractable analytically (Young 1976).  
An analytical formula of the 3-dimensional light distributions whose projected line profile resembles the $R^{1/n}$ S\'ersic law or the $R^{1/4}$ de Vaucouleaurs (1948) law is of great interest (Jaffe 1983; Hernquist 1990; Ciotti 1991; Dehnen 1993; Trijillo 2002). The Dehnen (1993) model is about spherically symmetric light distributions,
\begin{equation} 
\rho (r) = \frac {b}{r^\gamma }\frac {1}{ (r+a)^{4-\gamma }}
\end{equation}
where $a ,b$ are constants.

Another aspect of elliptical galaxy study is their shapes. Elliptical galaxies gain their name by their projected apparent shapes.  
However, their 3-dimensional shapes, axisymmetric or triaxial, are hard to infer (Binggeli 1980; Benacchio \& Galletta 1980; Binney \& de Vacouleurs 1981; Lambas, Maddox, \& Loveday 1992; Tenjes {\it et al.} 1993; Ryden 1996).
The assumption of triaxial shapes is extensively used. Stark (1977) proposed a model of triaxial ellipsoid isophotes of constant axis ratios whose projected isophote curves are exact ellipticals of constant axis ratios too.
The isophote curves of real images, however, have radially varying axis ratios. 
Carter (1978)'s investigation indicates that ellipticity of the isophotes is an increasing or peaked function of radius. 
Analysis of model galaxies suggests a similar result (Madejsky \& Mollenhoff 1990; Ryden 1991). 
In addition, the isophote curves are not exact elliptical. They are sometimes boxy-shaped and sometimes discy-shaped (Binney \& Merrifield 1998). 
Because the model of 3-dimensional nested isophote ellipsoids
with varying axis ratios involves an infinite amount of freedom of degrees, its projection has to be divided into finite layers and roughly fitted to images. 

One goal of the main astrophysical extragalactic studies is to find the symmetry behind the exponential law of disc galaxies and the S\'ersic law of elliptical galaxies. With the motivation, I proposed to use orthogonal curvilinear coordinate lines to model galaxy patterns (He 2003).     
Let $\rho(x,y)$ denote spiral galaxy light distribution on 2-dimensional $x$-$y$ plane (spiral galaxy disc plane). 
The orthogonal curvilinear coordinate system $(\lambda ,\mu )$ is
\begin{equation} 
\begin{array}{l}
x=x(\lambda,\mu), \\
y=y(\lambda, \mu)
\end{array}
\end{equation}
where $(x,y)$ is the common Cartesian coordinates.
We always use logarithmic light distribution, 
\begin{equation} 
f(x,y)=\ln \rho(x,y).
\end{equation}
Its gradient is $\nabla f$ whose Cartesian components are $(f^\prime _x, f^\prime _y)$. However, its components associated with the local reference frames of the curvilinear coordinate system are denoted by $u(\lambda ,\mu ) $ and $v(\lambda ,\mu ) $ (in the positive directions of $d\lambda $ and $d\mu $ respectively). 
They are also the directional derivatives to $f(x,y)$ in the directions of the curvilinear coordinate lines.   
Our main idea is that $ u(\lambda ,\mu ) $  depends only on $\lambda  $, $u(\lambda ,\mu )  \equiv u( \lambda ) $ which is the directional derivative to $f(\lambda ,\mu )$ along the coordinate line $\mu =$ constant. Similarly,
$ v(\lambda ,\mu ) $  depends only on $\mu  $, $v(\lambda ,\mu )  \equiv v( \mu ) $ which is the directional derivative to $f(\lambda ,\mu )$ along the coordinate line $\lambda =$ constant. 
That is, spiral galaxy light distribution $f(x,y)(=\ln \rho(x,y))$ is such that its gradient vector field can be decomposed in some curvilinear coordinate systems with corresponding components depending on single coordinate variables. 
This is called the single-variable requirement, or the symmetry principle.  
The symmetry principle maps an orthogonal curvilinear coordinate system to a logarithmic light distribution.
He (2005a) found the symmetric generalization of the simple orthogonal coordinate system which is formed by the complex-plane transformation $Z=e^W$.
The corresponding light distribution turns out to be spiral galaxy disc. Its radial density profile is exponential which is consistent with the known exponential law of spiral galaxy surface brightness. 
In He (2005b), a ``disturbed'' form of the coordinate system  is used to model galactic bars.  
In the present paper, I generalize the idea to 3-dimensional light distributions and present a model for elliptical galaxy patterns. 
An orthogonal coordinate system is employed which is formed by the complex plane transformation $Z=1/W$. 
The resulting spatial (3-dimensional) light distribution is spherically symmetric and has infinite gradient at the centre, which is called spherical-nucleus solution and is used to model galaxy central areas. We can make changes of the coordinate system by cutting out some column areas of its definition domain, the areas containing the galaxy centre.   The resulting spatial (3-dimensional) light distributions are axisymmetric or triaxial and have zero gradient at the centre, which are called elliptical-shape solutions and are used to model galaxy global elliptical structures. 
The two kinds of logarithmic light distributions are added together to model the full elliptical galaxy patterns. The model is the generalization of Dehnen's model (2) whose projected profile resembles de Vaucouleurs' law.     
One of the elliptical-shape solutions permits realistic numerical calculation and 
is fitted to all R-band elliptical images from the Frei {\it et al. }(1996)'s galaxy sample.
The model fits real galaxy shapes satisfactorily.
This suggests that elliptical galaxy patterns can be represented in terms of a few basic parameters. 

Section 2 is devoted to constructing the spherical-nucleus solution.  
Section 3 presents axisymmetric and triaxial elliptical-shape solutions. 
The two types of solutions are combined to give a full model of elliptical galaxy patterns in section 4. A simple discussion of the ellipticity of the projected isophotes is given in the same section.
Section 5 is the fitting steps of the model to real galaxy images.
Section 6 is simple discussion.

\section{ A Model of Galaxy Central Areas: the Spherical-nucleus Solution  }
\begin{enumerate} 
\item  {\it Orthogonal coordinates.}
I propose a spherically symmetric solution of 3-dimensional light distributions. Spherically symmetric patterns are also axisymmetric:
\begin{equation} 
f(r) \equiv f(x,y,z)\equiv f(R, z),
\end{equation}
where $R=\sqrt{x^2+y^2}$ and the $z$-axis is the axis about which the 3-dimensional light distribution is rotationally symmetric, i.\,e. axisymmetric. Therefore,  we first introduce the standard description of orthogonal curvilinear systems defined on a 2-dimensional plane $(R,z)$ which is any plane containing the $z$-axis (see Fig.\,1). A curvilinear coordinate system defined on the plane labels each point $(R,z)$ with an ordered set of two real numbers $(\lambda, \mu )$ over a region $S$ on $(\lambda , \mu )$ plane, 
\begin{equation} 
\begin{array}{l}
R=R(\lambda,\mu), \\
z=z(\lambda, \mu).
\end{array}
\end{equation}
The positive increment $d \lambda >0$ defines the positive direction of the coordinate line $\mu = $ constant. Similarly for the coordinate line $\lambda = $ constant.
In terms of curvilinear coordinates $\lambda = \tau ^1, \mu =\tau ^2 $, the element of distance $dl$ between two adjacent points $(R,z)\equiv (\tau ^1, \tau ^2)$ and  
$(R+dR, z+dz)\equiv (\tau ^1+d\tau ^1, \tau ^2+d\tau ^2)$ is given by the quadratic differential form (Korn \& Korn, 1968),
\begin{equation} 
dl^2=dR^2+dz^2=\sum^2_{i=1} \sum^2_{k=1}g_{ik}(\tau ^1, \tau ^2)d\tau ^id\tau ^k
\end{equation}
with
\begin{equation} 
g_{ik} = \frac {\partial R}{\partial \tau ^i}\frac {\partial R}{\partial \tau ^k}+\frac {\partial z}{\partial \tau ^i}\frac {\partial z}{\partial \tau ^k}, \; i,k=1,2.
\end{equation} 
Orthogonal coordinate system is such that the array $(g_{ik})$ is diagonal,
\begin{equation} 
g_{12}=g_{21}=\frac {\partial R}{\partial \lambda }\frac {\partial R}{\partial \mu }+
\frac {\partial z}{\partial \lambda }\frac {\partial z}{\partial \mu} \equiv 0.\\
\end{equation}

\begin{figure}
 \mbox{} \vspace{8.9cm} \includegraphics{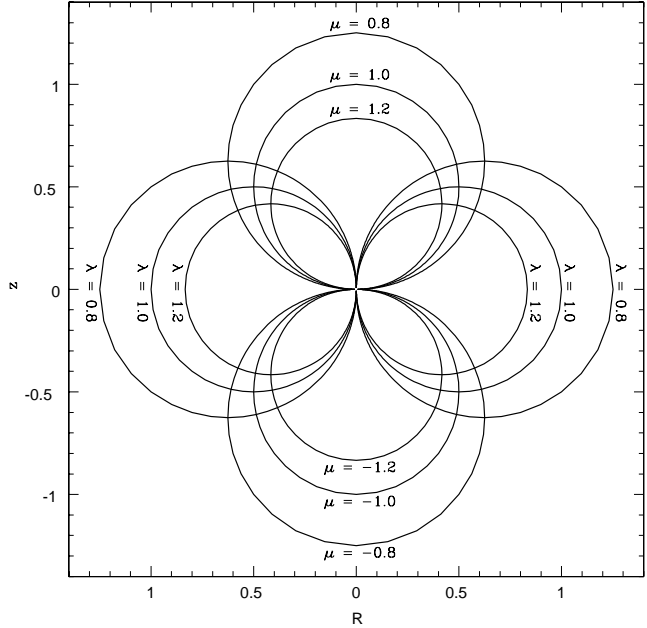} \caption[]{The orthogonal 4-circle coordinate system. The $(R,z)$ plane is any one in 3-dimensional space which contains the Cartesian $z$-axis. Therefore, $R=\surd(x^2+y^2)$.   }
\end{figure}

We frequently need to trace quantities along curvilinear coordinate lines especially when we draw a pattern. Here we consider arc-lengths measured along curvilinear coordinate lines. We denote the 
arc-length along the line $\mu =$ constant by $s(\lambda , \mu )$ and along the line 
$\lambda =$ constant by $t(\lambda , \mu )$. According to the formula (7), they can be calculated as follows,
\begin{equation} 
\begin{array}{l}
ds= \sqrt{g_{11}} d\lambda = \sqrt {R_\lambda ^{\prime 2} +  z_\lambda ^{\prime 2} } d\lambda =P d\lambda , \\
dt=\sqrt{g_{22}} d\mu =\sqrt {R_\mu ^{\prime 2} +  z_\mu ^{\prime 2} } 
d\mu =Q d\mu  
\end{array}
\end{equation}
where we have introduced notations for arc-length derivatives,
\begin{equation} 
\begin{array}{l}
P=s_\lambda ^\prime = \sqrt {R_\lambda ^{\prime 2} +  z_\lambda ^{\prime 2} }= \sqrt{g_{11}} ,  \\
Q=t_\mu ^\prime =\sqrt {R_\mu ^{\prime 2} +  z_\mu ^{\prime 2} }=\sqrt{g_{22}}  . 
\end{array}
\end{equation}

\item  {\it Complex Plane Reciprocal Transformation and Symmetry Principle. }
Now we present the coordinate system on which  all the models proposed in the paper are based, 

\begin{equation} 
\begin{array}{l}
R = \lambda /(\lambda ^2 +\mu ^2) ,   \\
z = \mu /(\lambda ^2 +\mu ^2), \\
0<\lambda <+\infty , \; -\infty <\mu <+\infty . 
\end{array}
\end{equation}
The orthogonal coordinate system can be produced by the complex-plane transformation $Z=1/W$
where $Z=R+{\rm i}z$ and  $W=\lambda -{\rm i}\mu $.
The corresponding arc-length derivatives $P,Q$ are
\begin{equation} 
\begin{array}{l}
P(\lambda,\mu) = s_\lambda ^\prime =1/(\lambda ^2+\mu ^2),\\
Q(\lambda,\mu) = t_\mu ^\prime = 1/(\lambda ^2+\mu ^2)\equiv P.\\
\end{array}
\end{equation}
The inverse formulas of the coordinate transformation are
\begin{equation} 
\begin{array}{l}
\lambda = R /(R ^2 +z ^2) ,   \\
\mu = z /(R ^2 +z ^2), \\
0<R <+\infty , \; -\infty <z <+\infty . 
\end{array}
\end{equation}
Now we look for the logarithmic galaxy light distribution $f(R,z)$ defined on the $(R,z)$ plane. The curl of the gradient vector field of the light distribution $f(R,z)$ must be zero. We work on curvilinear coordinates and we know that the components of the gradient associated with the coordinate are $u$ and
$v $. The zero-curl condition is given in standard mathematical books, e.g.  Korn \& Korn (1968),
\begin{equation} 
\frac {\partial }{\partial \mu }(u(\lambda ,\mu) P)- 
\frac {\partial }{\partial \lambda }(v(\lambda ,\mu) Q)=0 
\end{equation}
where $P=\sqrt{g_{11}} $,  $Q=\sqrt{g_{22}} $.  Because we require $u,v$ be functions of single variables $\lambda $ and $\mu $ respectively (symmetry principle), the above general zero-curl condition changes into
(He, 2005a),
\begin{equation} 
u(\lambda ) P_\mu -v(\mu )Q_\lambda =0.
\end{equation}

Now we need to apply the equation to our circle-shaped coordinate system. Substitution of the formulas (13) into above equation, we have
the gradient vector field $(u,v)$ for our circle-shaped coordinate, 
\begin{equation} 
u(\lambda ) = h\lambda ,v(\mu) =h\mu
\end{equation}
where $h$ is constant. 
The galaxy light distributions along all orthogonal coordinate lines can be found by performing path integrations of the following formulas 
\begin{equation} 
\begin{array}{l}
df=u ds = u(\lambda) P d \lambda ,\\
df=v dt = v(\mu ) Q d \mu 
\end{array}
\end{equation}
along the coordinate lines $\mu =$ constant and $\lambda =$ constant respectively.
Choosing the galaxy origin to be $R=0,z=0$ and performing the path integration, we can have our light distribution on the $(R,z)$ plane,
$ f = (h/2)\ln (\lambda ^2+\mu ^2)$.
We can prove the formula by verifying the relation between the physical components of its gradient, $u, v$, and the contravariant components, $\partial f/\partial \lambda , \partial f/\partial \mu $. The relation is, $ u= (\partial f/\partial \lambda )/P, v= (\partial f/\partial \mu )/Q$.
Using the inverse formulas of the coordinate transformation (14), we finally have our logarithmic and direct light distributions respectively, 
\begin{equation}
\begin{array}{l}
f(R,z) = (h/2)\ln (\lambda ^2+\mu ^2) =(h/2)\ln \frac{1}{R ^2+z ^2},\\
\rho (R,z)=\rho_0 (\frac{1}{R^2+z^2})^{h/2}, \\
0<R <+\infty , \; -\infty <z <+\infty . 
\end{array}
\end{equation}
where $\rho _0$ is another constant.
We choose $h>0$ because light $\rho =\rho _0\exp f \rightarrow 0$ when 
$R, z \rightarrow +\infty $. 

\item  {\bf A Model of Galaxy Central Areas: the Spherical-nucleus Solution. }
We rotate the above $(R,z)$ plane about the $z$-axis (i.e. the line $R=0$) to achieve
our 3-dimensional galaxy light distribution which is rotationally symmetric, i.\,e. axisymmetric. The resulting light distribution is 
\begin{equation} 
\begin{array}{l}
f_n (x,y,z) = (h/2)\ln \frac{1} {x^2+y^2+z^2},\\
\rho _n(x,y,z)=\rho_0 (\frac{1}{x^2+y^2+z^2})^{h/2} 
     =\rho_0 \frac{1}{r^{h} }\\
 -\infty <x,y,z <+\infty . 
\end{array}
\end{equation}
We see that it is not only rotationally symmetric but also spherically symmetric. 
The light flux at the galaxy centre is $+\infty $ and the gradient of the light flux at the same centre is $-\infty $. The solution is called spherical-nucleus solution and is used to model galaxy central areas. In fact, it corresponds to the first factor of the Dehnen model (2)
with $\gamma$ being replaced by $h$.
We can project the 3-dimensional distribution on any plane to see its surface brightness. We choose the  $x$-$y$ plane to be the sky plane.
The surface brightness is     
\begin{equation} 
\begin{array}{ll}
\Sigma (x,y) &= \int ^{+\infty }_{-\infty } \rho (x,y,z)dz \\
             &=2\rho_0\int ^{+\infty }_0 (\frac{1}{x^2+y^2+z^2})^{h/2} dz \\
             &=2\rho_0\int ^{+\infty }_0 (\frac{1}{R^2+z^2})^{h/2} dz \\
             &=\Sigma (R)\\
             &=\rho _0 \frac{\sqrt{\pi }}{R^{h-1}}\frac{\Gamma ((h-1)/2)}{\Gamma (h/2)}.
\end{array} 
\end{equation}
where Gamma function is used and 
the integral can not be carried out to be an elementary function of the parameter $h$. The surface brightness at the galaxy centre is $+\infty $ which is inconsistent with the S\'ersic law. 
\end{enumerate} 

\section{Models of Galaxy Global Patterns: the Elliptical-shape Solutions  }
In this section I present axisymmetric and triaxial models of elliptical galaxy patterns.
\begin{enumerate} 
\item  {\bf A model of 3-dimensional axisymmetric light distributions.}
The model is a corrected form of the above 3-dimensional light distribution (20), 
\begin{equation} 
\begin{array}{l}
f_{1s} (x,y,z) = (h/2)\ln \frac{1} {(R_0+R)^2+(z_0+|z|)^2},\\
\rho _{1s} (x,y,z)=\rho_{0} (\frac{1}{(R_0+R)^2+(z_0+|z|)^2})^{h/2} \\
0<R <+\infty , \; -\infty <z <+\infty 
\end{array}
\end{equation}
where $R_0 (>0)$ and  $z_0(>0)$ are constants.
This is equivalent to cutting out an infinite long cylinder of light distribution (of radius $R_0$) around the $z$-axis and cutting out an infinite plane layer of light distribution (of width $2z_0$) parallel to and centreing on $x$-$y$ plane from the spherically symmetric distribution (20).
The corresponding orthogonal coordinate system is demonstrated in Fig.\,2 which is the one in Fig.\,1 with corresponding parts being cut out.
The corresponding coordinate equation system is
\begin{equation} 
\begin{array}{l}
R = \lambda /(\lambda ^2 +\mu ^2)-R_0 ,   \\
|z| = |\mu | /(\lambda ^2 +\mu ^2)-z_0, \\
0<\lambda <R_0(R_0^2+z_0^2)/z_0^2 , \; 0<|\mu | <(R_0^2+z_0^2)/z_0 . 
\end{array}
\end{equation}
We can see that the 3-dimensional light distribution (22) is rotationally symmetric and its projection on any sky plane is apparently elliptical except on the $x$-$y$ plane.
 
\begin{figure}
 \mbox{} \vspace{8.9cm} \includegraphics{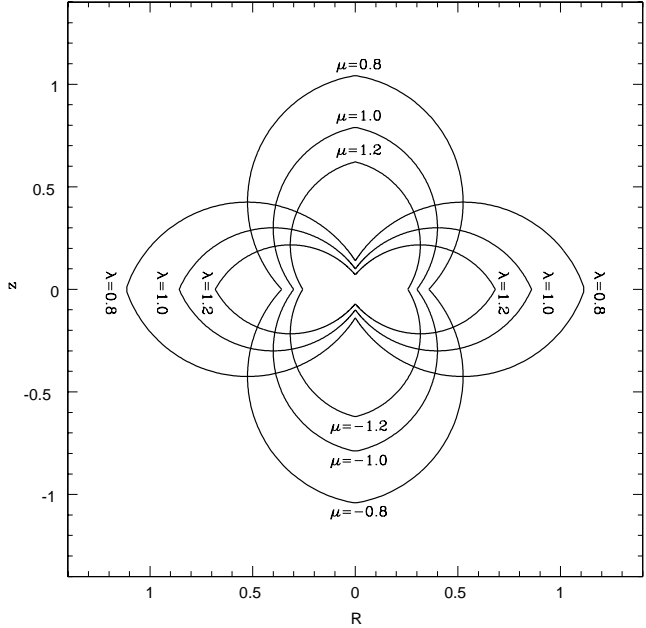} \caption[]{ The orthogonal coordinate system is obtained by simply cutting out two infinite columns of areas in Fig.\,1. The two columns are  parallel to and centreing on the Cartesian $R$-axis and $z$-axis respectively. The widths of the two columns are $2R_0$ and $2z_0$ respectively.   }
\end{figure}

\begin{figure}
 \mbox{} \vspace{8.9cm} \includegraphics{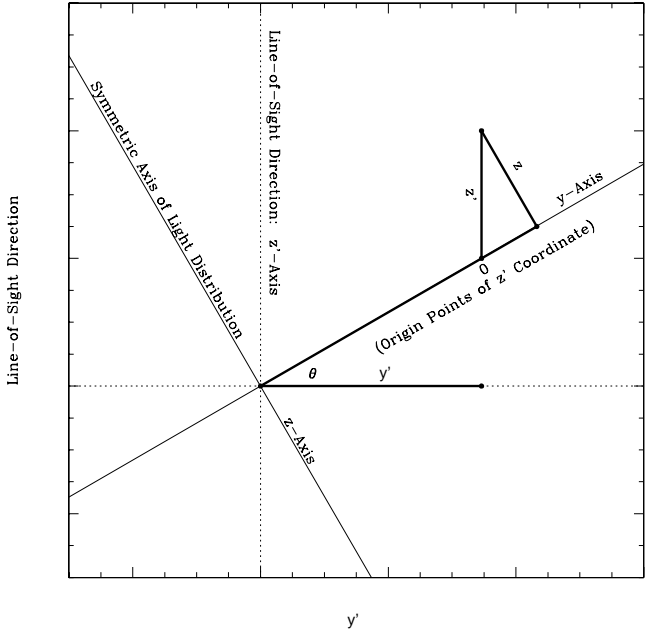} \caption[]{
we project the 3-dimensional light distribution on the sky plane which is a plane containing the $x$-axis and is obtained by rotating the $x$-$y$ plane about the $x$-axis with an angle $\theta _0$ (the inclination angle of the axisymmetric light distribution). }
\end{figure}

Now, we project the 3-dimensional light distribution on the sky plane which is a plane containing the $x$-axis and is obtained by rotating the $x$-$y$ plane about the $x$-axis with an angle $\theta _0$ (the inclination angle of the axisymmetric galaxy, see Fig.\,3). 
The coordinates $(x^\prime ,y^\prime )$ on the sky plane (see Fig.\,3) and the coordinate $z^\prime $ along the line-of-sight are related to the original coordinates $(x,y,z)$ as follows
\begin{equation} 
\begin{array}{l}
x=x^\prime , \\
y=y^\prime \sec \theta _0 + z^\prime \sin \theta _0, \\
z= z^{\prime}\cos \theta _0.  
\end{array} 
\end{equation}

The surface brightness on the sky plane after the projection
is given as the following integration,
\begin{equation} 
\begin{array}{l}
\Sigma (x^\prime ,y^\prime ) 
             = \int ^{+\infty }_{-\infty } \rho (x,y,z)dz^\prime \\
 =\rho_0\int ^{+\infty }_{-\infty } (\frac{1 }{(R_0+\sqrt{x^{ 2}
         +y^2})^2+(z_0+ |z|)^2 } )^{h/2}dz^\prime .
\end{array} 
\end{equation}
The $x^\prime $-axis is the major or minor axis of the 2-dimensional elliptical
light distributions $\Sigma (x^\prime ,y^\prime )$.
The integral can not be carried out to be an analytic formula. 
It can be numerically calculated.
However, it can be analytically solved in some special cases.

Firstly, take $h=3, \theta _0 =0$, and $z_0 =0 $. The corresponding surface brightness is
\begin{equation} 
\Sigma (x^\prime ,y^\prime ) = 2\rho_0/(R_0+R^\prime )^2 
\end{equation}
where $R^\prime =\sqrt{x^{\prime 2} +y^{\prime 2}}$. This is exactly the Hubble (1930) law.
Secondly, take $h=3, \theta _0 =0$, and $z_0 \not=0 $. The corresponding surface brightness is
\begin{equation} 
\Sigma (x^\prime ,y^\prime ) =
  \frac{2\rho _0}{(R_0+R^\prime )^2}(1-\frac{z_0}{\surd(z_0^2+ (R_0+R^\prime )^2)})
\end{equation}
which is a generalized Hubble law. The general case $(h>3)$ requires numerical calculation.
However, the meaning of the parameters is simple.
The model of axisymmetric light distributions (22) involves four parameters $ \rho_0 ,h, R_0, z_0 $.
The central light density of the light distribution is proportional to $\rho _0$ while $h$ represents galaxy light concentration and all other parameters ($R_0, z_0$ and $x_0, y_0$ in the following part (v)) affect the elliptical shapes of the 3-dimensional light distributions.  
The projection of the light distribution on to the sky plane (the integration (25)) can not be expressed by the $\Gamma $ function of $h$ as we did in (21).
The total luminosity, i.\,e. the summation of the light distribution through whole space is finite when $h>3$. 

\item{\bf The 3-parameter axisymmetric model. }
Our model of axisymmetric light distributions (22) involves 4 parameters ($ \rho_0 ,h, R_0, z_0 $).
Its projection on to the sky plane (the integration (25)) can not be expressed by the $\Gamma $ function of $h$ as we did in (21).
However, if we choose $R_0$ to be zero, our 4-parameter model changes into a 3-parameter model ($ \rho_0 ,h, z_0 $),
\begin{equation} 
\rho _{2s}(x ,y, z ) 
 =\rho_0 (\frac{1 }{x^2+y^2+(z_0+ |z|)^2 } )^{h/2}      
\end{equation}
and its projection on to the sky plane can be evaluated by the
$\Gamma $ function of $h$ and the integrals on finite intervals,
\begin{equation} 
\begin{array}{l}
\Sigma (x^\prime ,y^\prime )
 =\rho_0\int ^{+\infty }_{-\infty } (\frac{1 }{x^{2}
         +y^2+(z_0+ |z|)^2 } )^{h/2}dz^\prime   \\
 = \rho_0(\int ^{+\infty }_0 (\frac{1 }{x^{2}
         +y^2+(z_0+ z)^2 } )^{h/2}dz^\prime  \\
  +\int^0 _{-\infty } (\frac{1 }{x^{2}  
         +y^2+(z_0-z)^2 } )^{h/2}dz^\prime  ) \\
 = \rho_0\int ^{+\infty }_0 ((\frac{1 }
   {R_+^{ 2}+(Z_++ z^\prime )^2 } )^{h/2}  
   +(\frac{1 }{R_-^{ 2}+(Z_-+ z^\prime )^2 } )^{h/2} )dz^\prime \\
 = \rho_0(\frac{1 }{R^{h-1}_+}\int ^{+\infty }_{Z_+/R_+}(\frac{1 }
   {1+s^2 } )^{h/2}ds\\
    + \frac{1 }{R^{h-1}_-}\int ^{+\infty }_{Z_-/R_-}(\frac{1 }
   {1+s^2 } )^{h/2}ds )
\end{array} 
\end{equation}
where
\begin{equation} 
\begin{array}{l}
R_{\pm}=\sqrt{x^{\prime 2}
         +(y^\prime \mp z_0\sin\theta _0)^2 }\\
Z_{\pm}=\pm y^\prime \tan\theta _0 +z_0\cos\theta _0    
\end{array} 
\end{equation}
are not dependent on the integration variable $z^\prime $.
Finally, our 3 parameter model gives light distributions on the sky plane 
as follows,
\begin{equation} 
\begin{array}{l}
\Sigma (x^\prime ,y^\prime )= \rho_0( \\
\frac{1 }{R^{h-1}_+}(\frac{\sqrt{\pi }}{2}\frac{\Gamma ((h-1)/2)}{\Gamma (h/2)} +   \int ^0_{Z_+/R_+}(\frac{1 }
   {1+s^2 } )^{h/2}ds)\\
    + \frac{1 }{R^{h-1}_-}(\frac{\sqrt{\pi }}{2}\frac{\Gamma ((h-1)/2)}{\Gamma (h/2)} +
\int ^0_{Z_-/R_-}(\frac{1 }
   {1+s^2 } )^{h/2}ds )). 
\end{array} 
\end{equation}
This is exactly the model which is used in section 5  to be fitted to real elliptical galaxy images. 

\item {\it Orthogonal coordinates and zero-curl equations in the case of 3-dimensional space.}
The curvilinear coordinate system in 3-dimensional space is 
\begin{equation} 
\begin{array}{l}
x=x(\lambda,\mu, \nu), \\
y=y(\lambda,\mu, \nu), \\
z=z(\lambda,\mu, \nu).
\end{array}
\end{equation}
The arc-length derivatives are,
\begin{equation} 
\begin{array}{l}
P=s_\lambda ^\prime = \surd (x_\lambda ^{\prime 2} + y_\lambda ^{\prime 2} +   z_\lambda ^{\prime 2} )= \sqrt{g_{11}} ,  \\
Q=t_\mu ^\prime = \surd (x_\mu ^{\prime 2} + y_\mu ^{\prime 2} +   z_\mu ^{\prime 2} )= \sqrt{g_{22}} ,  \\
G=\surd (x_\nu ^{\prime 2} + y_\nu ^{\prime 2} +   z_\nu ^{\prime 2} )= \sqrt{g_{33}} .
\end{array}
\end{equation}
We work on curvilinear coordinate system and the components of the gradient vector of the logarithmic light distribution $f(x,y,z)$ associated with the coordinate system are $u, v$ and $w $. Because they are functions of the single variables $\lambda , \mu , \nu $ respectively, the zero-curl condition consists of the following equations,
\begin{equation}
\begin{array}{l}
u(\lambda ) P_\mu -v(\mu )Q_\lambda =0,\\
u(\lambda ) P_\nu -w(\nu )G_\lambda =0,\\
v(\mu ) Q_\nu -w(\nu )G_\mu =0.
\end{array}
\end{equation}

Now we present the 6-sphere coordinate $(\lambda ,\mu, \nu )$ which is the generalization of the plane 4-circle coordinate (formulas (12) and Fig.\,1) to the case of 3-dimensional space, 
\begin{equation} 
\begin{array}{l}
x = \lambda /(\lambda ^2 +\mu ^2 +\nu ^2) ,   \\
y = \mu /(\lambda ^2 +\mu ^2 +\nu ^2) ,   \\
z = \nu /(\lambda ^2 +\mu ^2 +\nu ^2) ,   \\
-\infty <\lambda ,\mu ,\nu <+\infty . 
\end{array}
\end{equation}
In Section 2, the plane 4-circle coordinate lines are rotated about the $z$-axis. The resulting 3-dimensional coordinate surfaces consist of one family of 2-spheres, one family of toroids, and the other family of all planes containing the $z$-axis. Here the coordinate system has triaxial symmetry and the coordinate surfaces consist of three families of 2-spheres. Therefore, the coordinate is called a 6-sphere coordinate (Moon \& Spencer, 1961).  
The corresponding arc-length derivatives $P,Q, G$ are
\begin{equation} 
\begin{array}{ll}
P(\lambda,\mu, \nu)& = 1/(\lambda ^2+\mu ^2 +\nu ^2)\\
                   &\equiv Q(\lambda,\mu, \nu) \equiv G(\lambda,\mu, \nu) . 
\end{array}
\end{equation}
The inverse formulas of the coordinate transformation are
\begin{equation} 
\begin{array}{l}
\lambda = x /(x ^2 +y^2+z ^2) ,   \\
\mu = y /(x ^2 +y^2+z ^2) ,   \\
\nu = z /(x ^2 +y^2+z ^2) , \\
-\infty <x, y, z <+\infty . 
\end{array}
\end{equation}
The zero-curl equation system (34) determines the gradient vector field $(u,v,w)$
of our 3-dimensional logarithmic light distribution, 
\begin{equation} 
u(\lambda ) = h\lambda ,v(\mu) =h\mu , w(\nu)=h\nu
\end{equation}
where $h$ is a constant. 
The galaxy light distributions along all orthogonal coordinate lines can be found by performing path integrations of the gradient components,
\begin{equation}
\begin{array}{l}
f_n = (h/2)\ln (\lambda ^2+\mu ^2 +\nu ^2) =(h/2)\ln \frac{1}{x ^2 +y^2+z ^2},\\
\rho _{n}(x,y,z)=\rho_0 (\frac{1}{x^2+y^2+z^2})^{h/2}  =\rho_0 \frac{1}{r^{h} }\\
-\infty <x, y, z <+\infty .
\end{array}
\end{equation}
We choose $h>0$ because light $\rho =\rho _0\exp f \rightarrow 0$ when 
$x,y, z \rightarrow +\infty $. 
The light distribution turns out to be the spherical-nucleus solution. 
However, we can achieve triaxial and other light distributions based on the result, using the cut-out method. 

\item {\bf A model of spherically symmetric light distribution with zero gradient at the galaxy centre.}
We can cut out a sphere (radius $a$) of light distribution centred at the galaxy nucleus. The resulting light distribution is 
\begin{equation} 
\begin{array}{l}
f_{3s} (x,y,z) = (h/2)\ln \frac{1} {(r+a)^2},\\
\rho _{3s}(x,y,z)  =\rho_0 \frac{1}{(r+a)^{h} },\\
 0<r <+\infty 
\end{array}
\end{equation}
In fact, it corresponds to the second factor of the Dehnen model (2) with $4-\gamma$ being replaced by $h$.

\item {\bf A model of 3-dimensional triaxial light distributions.}
We can cut out three infinite plane layers of light distributions parallel to and centreing on the three Cartesian coordinate planes respectively from (39). The three planes are the $x$-$y$ plane, $y$-$z$ plane, and $z$-$x$ plane. The widths of the three layers are $2x_0, 2y_0, 2z_0$ respectively.  
The resulting logarithmic light distribution is 
\begin{equation} 
\begin{array}{l}
f_{4s} (x,y,z) = (h/2)\ln \frac{1} { (x_0+|x|)^2+(y_0+|y|)^2+ (z_0+|z|)^2},\\
\rho _{4s}(x,y,z)  =\rho_0 (\frac{1} { (x_0+|x|)^2+(y_0+|y|)^2+ (z_0+|z|)^2})^{h/2},\\
 -\infty <x, y, z <+\infty 
\end{array}
\end{equation}
where $x_0 (>0), y_0 (>0)$ and  $z_0(>0)$ are constants.

Now, we project the 3-dimensional light distribution on the sky plane to see the surface brightness.
Take the $x$-$y$ plane to be the sky plane as an example.
To achieve an analytic result, take $h =3$. The resulting surface brightness is
\begin{equation} 
\begin{array}{ll}
\Sigma (x,y )& =
  \frac{2\rho _0}{ (x_0+|x| )^2+(y_0+|y| )^2}\cdot  \\
  & (1-\frac{z_0}{\surd(z_0^2+(x_0+|x| )^2+(y_0+|y| )^2  )})
\end{array}
\end{equation}
which is similar to the formula of the circularly symmetric light distribution (27). But here the surface brightness is non-circularly symmetric even when the sky plane is the $x$-$y$ plane.
\end{enumerate}

\section{A Model of Full Galaxy Patterns and Some Symmetry Results.  }

\begin{enumerate} 

\item{\bf A Model of Full Galaxy Patterns. } 
To achieve a full account of the central areas as well as the global shapes, we need to add the logarithmic light distribution of the spherical-nucleus solution in section 2 to one of the logarithmic light distributions of elliptical-shape solutions in section 3. The summation of two logarithmic light distributions is equivalent to the multiplication of their direct light distributions, $\rho _n \rho _{is}$.
If, instead of multiplication, we add together the direct light distributions, $\rho _n + \rho _{is}$, then we would see that elliptical galaxies consist of two apparent components. Real galaxy images, however, do not show such apparent components.  
The Dehnen model is recovered 
\begin{equation} 
\rho (r) =\rho _n (r) \rho _{3s}(r ) 
         =\rho _0 \frac {1}{r^{h_1} }\frac {1}{ (r+a)^{h_2 }}
\end{equation}
if we choose $h_ 1=\gamma , h_ 2=4-\gamma  $. 
This gives a model of spherically symmetric patterns.
The projected radial profile of the model light distribution resembles the de Vaucouleurs law. Its detailed discussion is given in Dehnen (1993). All models $\rho _n \rho _{is}$, $i=1,2,3,4$, in the present paper have similar properties. Here, we pay special attention to the $\rho _n \rho _{2s}$ axisymmetric model  
\begin{equation} 
\begin{array}{ll}
\rho (x,y,z) &=\rho _n \rho _{2s} \\
         &=\rho _0 \frac {1}{r^{h_ 1} }\frac {1}{  (x^2+y^2+(z_0+ |z|)^2 )^{h_2/2}  }
\end{array}
\end{equation}
which gives account of both galaxy central area and elliptical shape.
The model is called full 3-parameter axisymmetric model. 
The projection of the light distribution on to the sky plane can not be evaluated by the
$\Gamma $ function of $h$ and the integrals on finite intervals as we did in (31) for the original 3-parameter model.

\item{\bf Ellipticity as a function of radius. } 
Firstly, let us study the ellipticity of the projected isophote curves of the original 3-parameter model $\rho _{2s} $.
The 3-dimensional light distribution of the model is the result of a spherically symmetric one with a layer of light centred on the $x$-$y$ plane being cut out. Therefore, if the sky is a plane containing the $z$-axis then the projected light distribution (surface brightness) on the sky plane has approximately elliptical isophote curves (discy-shaped) whose ellipticity is approximately 
\begin{equation} 
\epsilon (R) =\sqrt {1-R^2/(R+z_0)^2 }.
\end{equation}
The ellipticity decreases monotonically with radius $R$ from $\epsilon =1$ at the galaxy centre to  $\epsilon =0$ at $R= +\infty $. Actually, we see $ 1>\epsilon _1 >\epsilon (R) >\epsilon _2 >0 $ because $\rho _{2s}$ has a flat light distribution near the galaxy centre and the image is always taken within a finite area. 

Secondly, we study the full 3-parameter axisymmetric model, $\rho (x,y,z) =\rho _n \rho _{2s}$.
The ellipticity $\epsilon (R )$ is a peaked function of $R$ 
changing from $\epsilon =0$ at the galaxy centre to  $\epsilon =0$ at $R= +\infty $, because the spherical-nucleus solution has dominant logarithmic light density ($\approx +\infty$) near the galaxy centre while all elliptical-shape solutions have finite values (flat light distributions) near the galaxy centre. Actually, we see an increasing ellipticity function if $z_0$ is large enough, or a decreasing function if $z_0$ is small enough,  or a peaked or flat function if $z_0$ has intermediate values. These are consistent with Carter (1978)'s image analysis.
The ellipticity function and the isophote-curve shapes (discy or boxy) change with the orientation of the line of sight and with the models of elliptical-shape solutions.

\item{\bf Light-strength invariance. }
Note that there is always an arbitrary factor $\rho _0$ accompanying our model light distribution (see, e.\,g. (21) and (22)). This is because our models involve the logarithmic light distributions $f(x,y,z)$ and the symmetry principle deals with their gradients only and leaves an integration constant for  $f(x,y,z)$.
This is called light-strength invariance which indicates that bright galaxies can share the same patterns with dim galaxies. Because galaxy visible mass distributions are approximately proportional to their light distributions, $\rho (x,y,z)$  can be mass distribution or light distribution  in our model.  In our fitting program, however, the projection of $\rho (x,y,z)$ on to the sky plane is fitted to the light distributions on galaxy images. 

The satisfactory fitting of our model to real images (see section 5) suggests that elliptical galaxy patterns can be represented in terms of a few basic parameters given in the model. Real galaxies are further constrained by their physical process. For example, the shape of the radial profile and the shape of isophote curves change with luminosity of the objects. Our mathematical model gives larger freedom of degrees and bright galaxies are allowed to share the same patterns with dim galaxies.

\item{\bf Scale invariance and the unit of lengths.}
We can multiply a constant $ \Lambda $ to the coordinate equations (12) to form a new coordinate system (scale transformation),
\begin{equation} 
\begin{array}{l}
\bar{R} =\Lambda \lambda /(\lambda ^2 +\mu ^2) ,   \\
\bar{z} =\Lambda \mu /(\lambda ^2 +\mu ^2) . 
\end{array}
\end{equation}
Then we have $\bar{P}=\Lambda P, \bar{Q}=\Lambda Q, \bar{u}=u/\Lambda $, and $ \bar{v}=v/\Lambda $.
The light distribution determined by the new coordinate system is
\begin{equation} 
\bar{\rho }(\bar{R}, \bar{z}) = \rho (R,z).
\end{equation}
This is called scale invariance. The scale transformation is equivalent to choosing different units for the Cartesian coordinates $R,z$.  The scale invariance says that our method of generating galaxy patterns does not depend on the choice of length units.
\end{enumerate}

\section{ Fitting the 3-Parameter Axisymmetric Model to Elliptical Galaxy Images }

Only the projections of the 3-parameter axisymmetric model $\rho _{2s}$ and the triaxial model $\rho _{4s}$ are found to be  
expressions of $\Gamma $ functions and integrals on finite intervals.
The projections of other models need further investigation.
The direct numerical calculation of the integrals on infinite intervals takes significant computer time to obtain a reasonable result.

In the present paper we fit the 3 parameter axisymmetric model (31) to real elliptical galaxy images and we find that the fitting is satisfactory except the areas near the galaxy centres. To achieve satisfactory fitting of full range of images, we need to apply the full model introduced in section 4 whose projection, however, is not an expression of $\Gamma $ functions and integrals on finite intervals and needs deep investigation.

\begin{figure}
 \mbox{} \vspace{8.9cm} \includegraphics{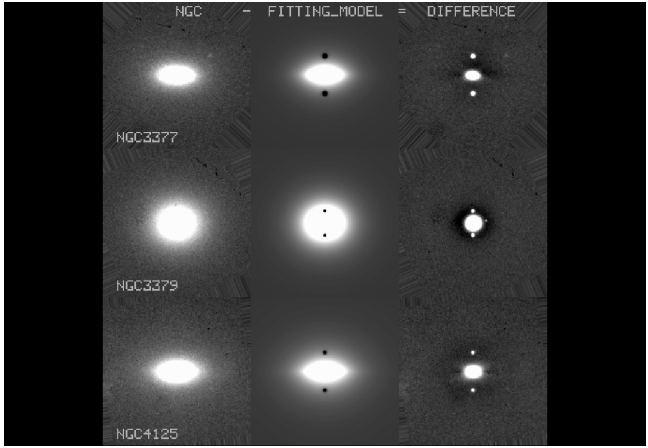} \caption[]{
The Frei {\it et al.}(1996) R-band images, NGC 3377, 3379, 4125, minus our theoretical fitting patterns respectively result in the residual images. Explanation of the spots on the fitting images are given in the Fitting Step 3.
}
\end{figure}

\begin{figure}
 \mbox{} \vspace{8.9cm} \includegraphics{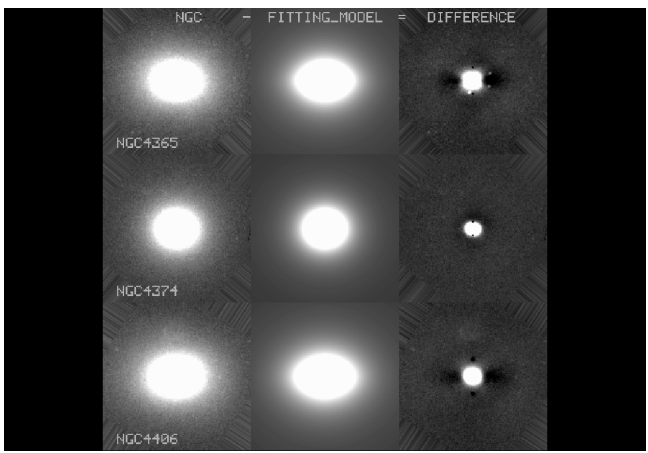} \caption[]{
The Frei {\it et al.}(1996) R-band images, NGC 4365, 4374, 4406, minus our theoretical fitting patterns respectively result in the residual images. Explanation of the spots on the fitting images are given in the Fitting Step 3.
}
\end{figure}

  {\it Fitting Step 1: Rotating the real elliptical galaxy image.}
We need to fit the expression $\Sigma (x^\prime ,y^\prime )$ (31) to real elliptical galaxy images. Because we cut out an infinite layer of light distribution centreing on the $x$-$y$ plane and we rotate the light distribution about the $x$-axis, the $x^\prime $-axis must be the major axis of the projected 2-dimensional elliptical
light distributions $\Sigma (x^\prime ,y^\prime )$ (see (24) and Fig.\,3). 
Therefore, we need to find the galaxy centre of the image and rotate the image about the centre so that the major axis of the elliptical galaxy image is in the horizontal direction as the ones of the model patterns are.

The sky level needs to be subtracted.
The resulting light distribution of the galaxy image is denoted by
\begin{equation} 
\hat \Sigma (i ,j )
\end{equation}
where $(i,j)$ are the image pixel coordinates.

 {\it Fitting Step 2: One more fitting parameter $l$: the side length of the square variance domain of $(x^\prime ,y^\prime )$ of the model formula.  }
The distance $d$ to us of a galaxy can be obtained by the Hubble law $d=cz/H$, given the Hubble constant $H$ (e.\,g., $H$=75 km/(s$\cdot$Mpc)) and the redshift $z$ of the galaxy. Therefore, the physical position $(\hat x,\hat y)$ on an elliptical  galaxy image is known because the angles subtended by the positions against us are measurable by the observer. 
Therefore, real galaxy light density $\hat \Sigma (\hat x, \hat y)$ can be considered the function of real physical position $(\hat x, \hat y)$.
The analytic formula (31), however, is a specific mathematical function $\Sigma $ of independent variables $x^\prime , y^\prime $ which can be denoted by whatever letters and have no physical meaning.
However, if real galaxy light distributions $\hat \Sigma (\hat x, \hat y)$ of all galaxies can be expressed by a single universal function then we can say that $x^\prime , y^\prime $ are proportional to physical position variables if we can testify that the formula (31) approaches to be the function.
The test consists of two steps.    
The real image is obtained from a finite square range of $\hat x, \hat y$ on a galaxy and the side length of the range, $\hat l$, is called the physical size of the data. The formula (31) is defined on $-\infty <x^\prime , y^\prime <+\infty $ and 
we need to choose a finite square range of $x^\prime , y^\prime $ for the theoretical formula, that is, we need to choose a part of the theoretical light distribution (31) which is compared to the digital image data.
If the theoretical formula approaches to be the exact expression of all galaxy light distributions then there must exist the part of the theoretical light distribution (31) which best fits the real galaxy data, along with the fitted values of all other parameters the formula involves. 
The side length $l$ of the fitted square range of $x^\prime , y^\prime $ is called the fitted size of the data.
We calculate the ratio of the fitted size to the physical size of the data, $l/\hat l$. We do this for a sample of galaxies. This is the first step.
The second step is very simple. We see if all the ratios are approximately identical. If they are, we can say that the galaxy light distributions $\hat \Sigma (\hat x, \hat y)$ of all galaxies can be expressed by a single universal function and the formula (31) approaches to be the function and the  mathematical variables $x^\prime , y^\prime $ have physical position meaning. 
A physical unit is qualified for the variables $x^\prime , y^\prime $, denoted by Ch. Its equivalent value to the meter or pc is given by the identical ratio.
Testing the identical ratio is equivalent to testing the proportionality of galaxy redshifts $z$ to the fitted lengths $l_0$ of one arcsecond distances on  the galaxy images. One arcsec distance on a galaxy image means that the corresponding physical distance on the galaxy makes an angle of one second at us (the observer). The fitted size $l$ divided by the total arcseconds of one side of the  corresponding galaxy image is the fitted arcsecond length $l_0$.
Whenever we are given a mathematical formula we should not lose the chance to test if it approaches to be the universal physical one. 

The first step of the test is explained on more detail.
As explained in the previous paragraph, the length of the square range of $x^\prime , y^\prime $ needs to be considered a fitting parameter (we do not worry about the centre position of the range because 
the centre of the model light distribution is exactly the galaxy centre).
We try different values of the length $l$.
For each value, the square domain is divided into $M \times M$ small squares whose length is $l/M$. 
The value of the theoretical light at the $i\times j$-th square, $\Sigma (i,j)$, is compared with the image light value  at the $i\times j$-th pixel, $\hat \Sigma (i,j)$.
The total relative error is 
\begin{equation} 
err =\sum_{i,j=1}^{M}| \hat \Sigma (i,j)- \Sigma(i,j)|/\hat \Sigma (i,j).
\end{equation}
Note that we take the relative error not the absolute error because both the pixel light values and the signal-to-noise ratios are very large near galaxy centres. Even the best fitting function, the S\'ersic law, leaves large absolute errors near the galaxy centres. To reduce the effects of the galaxy central areas, I choose relative error for the fitting.
The set of fitted values of the above scale $l$ and  all other parameters of the model formula makes the minimum $err$.  
 
 {\it  Fitting Step 3: Run computer program to find the fitted values of all parameters. }
I wrote a computer program for the fitting, that is, I try all possible values of the set of parameters and calculate the corresponding $\Sigma(i,j)$ (31) and the error (49).
 Finally I find the fitted values of the parameters which make the minimum $err$ (see the result of the second half of the Table 1). The corresponding fitted elliptical pattern is shown in  the middle columns of Figures 4 and 5). Note the two symmetric small spots on the central vertical direction of each fitted pattern in the Figures. This is due to almost zero values of $R_+$ and $R_-$ at or near the two points (see (30)).
Therefore, the integrals in (31) are on almost infinite intervals and their numerical calculations
result in large errors by the computer which runs over a limited time.
Now we know that it is important that we have an analytic model whose calculation can be taken on computers in a reasonable amount of time.

\begin{table}
\caption{The Fitted  Parameter Values for the S\'ersic Law and the 3-Parameter Model}
\begin{tabular}{lrrrrr} \hline

NGC &redshift& $l_0$  &$\sigma_s$ &$n$ &$\sigma _0$  \\
    & & [/arcsec] &   &  &  \\
\hline
3377 & 0.0022 & 0.040  & 72782 & 3.2 & -5.41 \\
3379 & 0.0030 & 0.087  & 58621 & 3.0 & -3.88 \\
4125 & 0.0045 & 0.014  & 56853 & 3.5 & -6.75 \\
4365 & 0.0042 & 0.038 & 63469 & 3.2 & -4.65  \\

4374 & 0.0035 & 0.050  & 51938 & 3.0 & -4.34 \\
4406 & -0.0008 & 0.017  & 47698 & 3.3 & -5.58 \\
4472 & 0.0033 & 0.036  & 67274 & 3.0 & -4.34 \\
4621 & 0.0014 & 0.037  & 26526 & 2.7 & -4.35 \\
4636 & 0.0031 & 0.017  & 50065 & 3.6 & -5.58 \\
5322 & 0.0059 & 0.033  & 134785 & 4.0 & -6.42 \\
5813 & 0.0066 & 0.036  & 34088 & 3.0 & -4.34 \\
\hline
NGC & $l_0$  & $\rho_0$ &$h$ &$\theta ^o $ & $z_0$  \\
    & [Ch/arcsec] &  &  &  & [Ch]  \\
\hline
 3377& 0.15 & 462862& 4.90 & 79 & 5.29 \\
 3379& 0.13  & 615071 & 5.00 & 37 & 4.78 \\
 4125& 0.28  & 479334 & 4.06 & 83 & 9.51 \\
 4365& 0.29 & 146908 & 3.46 & 74 & 7.33 \\
 4374& 0.35  & 159427 & 3.50 & 38 & 6.52 \\
 4406& 0.09  & 14994 & 3.62 & 66 & 2.99 \\
 4472& 0.32  & 122068 & 3.24 & 66 & 5.71 \\
 4621& 0.12  & 241165 & 4.86 & 71 & 4.41 \\
 4636& 0.30  & 103418 & 3.27 & 66 & 7.91 \\
 5322& 0.53  & 990340 & 3.83 & 79 & 13.21 \\
 5813& 0.13  & 547757 & 4.75 & 65 & 5.05 \\
\hline
\end{tabular}
\end{table}

The second half of the Table presents the fitting result for all 11 R-band elliptical galaxy images of Frei {\it et al. }(1996)'s sample (the sample consists of nearby bright galaxies and the images are cleaned by the original authors). 
The second column of the half of the Table is the fitted lengths $l_0$ of one arcsec distances on the images. Fig.\,6 indicates that the fitted lengths $l_0$ are approximately proportional to galaxy redshifts. 
This is equivalent to say that the ratios of the fitted sizes to the physical sizes of the image data are approximately constant.
That is, the galaxy light distributions $\hat \Sigma (\hat x, \hat y)$ of all galaxies can be expressed by a single universal function and the formula (31) approaches to be the function and the mathematical variables $x^\prime , y^\prime $ have physical position meaning. 
A physical unit is qualified for the variables $x^\prime , y^\prime $, denoted by Ch. Its equivalent value to the meter or pc is given by the identical ratio as follows
\begin{equation} 
1[{\rm Ch}]= 0.00027[{\rm Mpc}]=270[{\rm pc}].
\end{equation}

\begin{figure}
 \mbox{} \vspace{8.9cm} \includegraphics{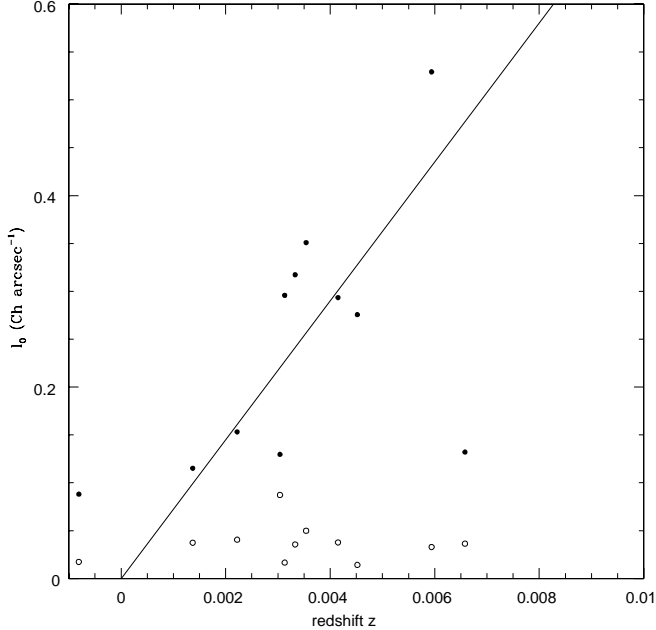} \caption[]{
The fitted lengths (filled circles) of one-arcsec distances on elliptical galaxy images  are approximately proportional to the galaxy redshists.  This is not true for the fitted lengths (open circles) based on one dimensional S\'ersic law.
}
\end{figure}

Figures 4 and 5 demonstrate the 2-dimensional fitting result. The fits along the galaxy major axes are shown in Fig.\,7. The observed profiles are indicated by the symbols.
  The solid lines are the fitting result taken from Figures 4 and 5. The dotted lines are the fitting result of the S\'ersic law  (1).
Similar to 2-dimensional model fitting,  when fitting  a one dimensional S\'ersic law I need to select a variance domain of $R$ for the function (1).  However,
I consider the length $l$ of the domain $(0, l)$ as a fitting parameter.
I try different values of the length $l$. For each value, the domain is divided into $M$ small intervals whose length is $l/M$. 
The value of the theoretical S\'ersic light at the $i$-th interval, $\Sigma _s (i)$, is compared with the image light value  at the $i$-th pixel along the major axis, $\hat \Sigma (i)$.
The total relative error is 
\begin{equation} 
err =\sum_{i=1}^{M}| \hat \Sigma (i)- \Sigma _s(i)|/\hat \Sigma (i).
\end{equation}
The set of fitted values of the length $l$ and  all other parameters of the S\'ersic law makes the minimum $err$.  In fig.\,7, the value at the upper right corner of each panel is the fitted value of $l$ for the corresponding galaxies.
The fitted size $l$ is divided by the total arcseconds of the major axis on the corresponding galaxy image. The result is $l_0$ which is the fitted length of one arcsec distance on the image. The fitted values of the parameters in the S\'ersic law together with $l_0$ are presented in the first half of Table 1.
The values of $l_0$ fitted by the S\'ersic law do not suggest a linear correlation with redshifts (open circles in Fig.\,6). One of the reasons is that an image is the projection of the corresponding 3-dimensional galaxy light distribution and the one dimensional S\'ersic law cannot give account of the galaxy shape.

\begin{figure}
 \mbox{} \vspace{8.9cm} \includegraphics{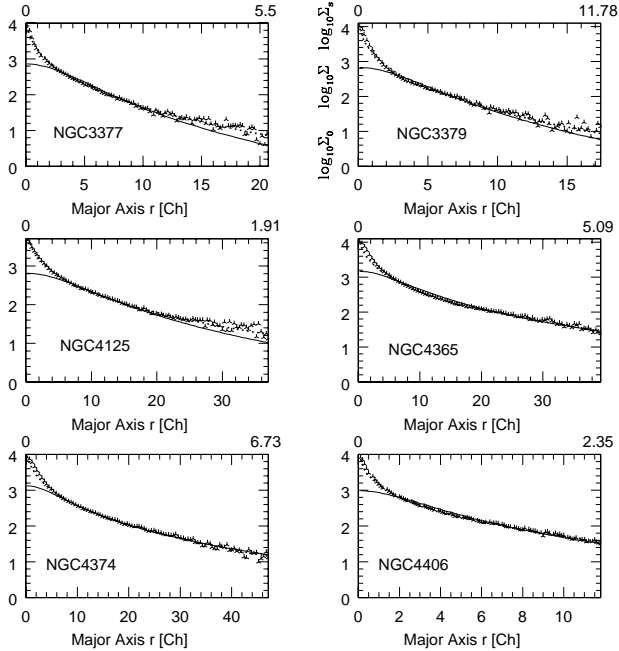} \caption[]{
The observed profiles are indicated by the symbols.
  The solid lines are the fitting result taken from Figures 4 and 5. The dotted lines are the fitting result of the S\'ersic law  (1).
 }
\end{figure}

\section{ Conclusion }

One goal of the main astrophysical extragalactic studies is to find the symmetry behind the exponential law of disc galaxies and the S\'ersic law of elliptical galaxies. With the motivation, I proposed to use orthogonal curvilinear coordinate systems and a symmetry principle to model galaxy patterns (He 2003).
The symmetry principle maps an orthogonal coordinate system to a light distribution pattern.
The coordinate system for elliptical galaxy patterns turns out to be the one which is formed by the complex-plane reciprocal transformation $Z=1/W$. 
The resulting spatial (3-dimensional) light distribution is spherically symmetric and has infinite gradient at its centre, which is used to model galaxy central area. By cutting out some column areas of its definition domain the coordinate system can also give axisymmetric or triaxial spatial light distributions.  
The two types of logarithmic light distributions are added together to model full elliptical galaxy patterns. The Dehnen model is recovered and generalized.
The isophote curves of the projected patterns can be discy or boxy and their ellipticity changes with radius. 
One of the elliptical-shape solutions permits realistic numerical calculations and is fitted to all R-band elliptical images from the Frei {\it et al. }(1996)'s galaxy sample. The fitting is satisfactory.
This suggests that elliptical galaxy patterns can be represented in terms of a few basic parameters.

The coordinate systems describing 2-dimensional disc galaxy patterns are based on the coordinate system which is formed by the complex-plane exponential transformation (He 2005b). He (2005b) found the evidence that the coordinate system can be considered a ``free fall'' coordinate space and a non-geometrized yet relativistic stellar dynamics was developed based on the principle of coordinate cancellation of gravity. A simple explanation of constant galactic rotation curves is given in He (2005b). The present paper indicates that elliptical galaxy patterns can be explained by the coordinate system (35). Whether the coordinate system cancels gravity in 3-dimensional space for elliptical galaxies needs deep investigation
and is left to be future work.

\section{Acknowledgments}
The author wishes to thank the anonymous referee for constructive guidance and Dr. B. Harms for providing the $\Gamma $ function expression in the formulas (21).

\end{document}